# Diabetic Retinopathy Detection using Ensemble Machine Learning

Israa Odeh
Department of Computer Science
PSUT
Amman, Jordan
isr20170294@std.psut.edu.jo

Mouhammd Alkasassbeh
Department of Computer Science
PSUT
Amman, Jordan
m.alkasassbeh@psut.edu.jo

Mohammad Alauthman
Department of Information Security
University of Petra
Amman, Jordan
mohammad.alauthman@uop.edu.jo

*Abstract*— Diabetic Retinopathy (DR) is among the world's leading vision loss causes in diabetic patients. DR is a microvascular disease that affects the eye retina, which causes vessel blockage and therefore cuts the main source of nutrition for the retina tissues. Treatment for this visual disorder is most effective when it is detected in its earliest stages, as severe DR can result in irreversible blindness. Nonetheless, DR identification requires the expertise of Ophthalmologists which is often expensive and time-consuming. Therefore, automatic detection systems were introduced aiming to facilitate the identification process, making it available globally in a time and cost-efficient manner. However, due to the limited reliable datasets and medical records for this particular eye disease, the obtained predictions' accuracies were relatively unsatisfying for eye specialists to rely on them as diagnostic systems. Thus, we explored an ensemble-based learning strategy, merging a substantial selection of well-known classification algorithms in one sophisticated diagnostic model. The proposed framework achieved the highest accuracy rates among all other common classification algorithms in the area. 4 subdatasets were generated to contain the top 5 and top 10 features of the Messidor dataset, selected by InfoGainEval. and WrapperSubsetEval., accuracies of 70.7% and 75.1% were achieved on the InfoGainEval. top 5 and original dataset respectively. The results imply the impressive performance of the subdataset, which significantly conduces to a less complex classification process when compared to the original complete Messidor dataset.

## I. INTRODUCTION

Diabetic Retinopathy is a diabetes complication that damages the light-sensitive retina tissues and blood vessels due to high blood sugar rates, macular changes such as yellowish spots, aneurysms (an increase of the microvascular thickness or "ballooning" in the retina), and hemorrhage (blood escaping from blood vessels) are considered the most common implications of DR.

Macular irregularities in diabetic patients were first detected in 1856 by Eduard Jaeger. However, those were not confirmed to be related to diabetes until 1872, when Jaeger first provided a histopathologic proof of "cystoid degeneration of the macula" in diabetic patients. Several studies were carried in the following years leading to the discovery of Proliferative Diabetic Retinopathy by Wilhelm Manz in 1876 [1].

As stated by Mayo Clinic [2], common symptoms of DR include spots in vision, blurred or fluctuated sight, color impairment, and in some severe cases, a complete vision loss in one or both eyes. In the long term, high blood sugar rates cause blockage in the microvessels of the retina, which are very important for nourishing the retina tissues, therefore, the eye attempts to grow new vessels to supply the retina with the needed nutrients and oxygen, however, these generated vessels are weak and likely to suffer blood leakage forming a hemorrhage in the retina. According to the severity of the detected symptoms, DR is graded into one of 3 stages; Mild, Moderate, and Proliferative DR (PDR).

In many cases, a fast clinical check and decision must be made for different reasons, such as a large number of patients in a specific facility, or an urgent and critical patient condition. Moreover, affordable treatment should be provided to all patients, however, in many developing countries, patients are not provided with adequate health care nor affordable treatment. Hence, many underprivileged patients are at a very high risk of losing their sight owing to the absence of reasonable healthcare. Consequently, various Artificial Intelligence algorithms were applied to produce efficient medical decision-making systems, such as in Expert Systems, Natural Language Processing (NLP) and other machine learning applications, leading to the first expert system specialized in medical practices called MYCIN, this rule-based prediction model was introduced in the early 1970s after almost 6 years of development at Stanford University, USA [3]. Many other Artificial Intelligence applications were employed in various healthcare sectors, like Radiology, Screening, and Disease Diagnosis. Several hospitals including Mayo Clinic, USA, and the National Health Service, UK have developed their own Intelligent systems [4,5], as well as Google [6] and IBM's [7] contributions to healthcare technology advancements.

In this research, we have developed a modern automatic detection model for Diabetic Retinopathy, concentrating on utilizing the most efficient ensemble of machine learning algorithms in order to obtain a highly accurate diagnosis. Moreover, in the present ensemble-based framework, as elucidated in the rest of this document, we have also considered achieving excellent precision while preserving an efficient performance with minimal time and storage costs.

The remainder of this paper is organized as follows; section 2 presents a literature review and a discussion of previously reported work on Diabetic Retinopathy automatic detection. In section 3, a detailed description of the proposed diagnostic model is provided, followed by the experimental dataset and methods used in this study. The final section concludes and summarizes the work along with the authors' opinion on future work and directions.

## II. Related Work

Numerous studies have been carried out on the automated identification of the DR, its reliability, efficiency and maintainability. In 2006, Jelinek et al. [8] proposed a DR detection system fully reliant on detecting red lesions in the retina using image processing and analysis. Followed by Abramoff et al. [9] in 2010, and Antal and Hajdu et al. [10] model in 2012 depending on the same primary lesion Jelinek used. However, the previously proposed systems haven't met the accuracy and sensitivity levels ophthalmologists required.

Further research was conducted throughout the following years. One of the ideas to improve the DR identification algorithms was to include more features that can be extracted from a retina funduscopy. For instance, image quality was proven to have a significant effect on the final prediction [11], retinal images with low resolution will probably increase the chance of developing FP and FN predictions.

Antal and Hajdu's [12] proposed system merges several comparison components used in previous works; Image quality, Lesion-specific components, Multi-scale AM/FM based feature extraction, Pre-screening and Anatomical components (Macula and Optic Disk detection) in an ensemble-based decision-making system, this was done by training several well-known classifiers along with energy functions and fusion strategies for ensemble selection. Basically, any classifier that produces a higher overall accuracy is included in the system, otherwise excluded. The authors recommended backward ensemble search methodology using accuracy and sensitivity energy functions, which first considers all classifiers are part of the ensemble, then each classifier is tested to be excluded only if elimination causes accuracy to increase. Antal and Hajdu's ensembled system achieved an outstanding accuracy of 90%, Sensitivity of 90%, and 91% Specificity in both disease and no-disease settings.

Gargeya and Leng's [13] proposed algorithm is mainly constructed of Deep convolutional neural networks (DCNNs), which consist of several filtering layers, each one produces the sum of both its outcome and the previous layer's results. In addition to sharp image preprocessing methodologies, putting great efforts on fundus image preprocessing enhanced the deep feature extraction methodologies used. Gargeya et al. depended on Sensitivity and Specificity measures to evaluate their system, in addition to the average Area Under the Receiver Operating Characteristic curve (AUROC). The system was tested to classify DR in 2 cases, the presence of the disease regardless of its stages against the healthy retinal state, as well as identifying the presence of the disease in very early stages with only few mild symptoms against its absence, as the results were visualized using a heatmap. On the MESSIDOR dataset, 93%, 87% and 0.94 of sensitivity, specificity and AUC were achieved by this algorithm for classifying no DR versus DR of any stage[14]. The overall predictive results when tested on the MESSIDOR-2 public dataset achieved slightly similar to Antal et al. model's results. Furthermore, the authors believe that was a consequence of the algorithm being unfamiliar with the MESSIDOR dataset as it was trained on the EyePACS public dataset (EyePACS LLC, Berkeley, CA) [15].

Training the model once on specific dataset images was the common approach used among previous developments, however, Zhou et al. [16] suggested using the concept of Multiple Instance Learning (MIL), dividing the training phase into two stages, single and multi-scale training, the first stage is to determine DR lesions for each image in one particular scale independently, and the latter to search for lesions in an image represented in multiple scales. The authors preprocessed images by normalizing, resizing and cropping them into a rectangular shape. Gaussian Smoothing Kernel was applied to adjust images' brightness, contrast, and color intensities. Images were resized to fit FOV with 384 pixel radius. Cropping was performed to ensure the elimination of bright borders of the image. In the single-scale learning, all images were scaled to r = 384-pixel. Suspicious patches were collected and tested individually to estimate their probability of being a DR lesion via the pre-trained AlexNet adapted in a CNN-based patch-level classifier. The estimations for each and all patches in the one image are fed into a global aggregation function, which determines the probability of the image (from which patches were collected) to represent a DR case using a developed DR map. In the second learning stage, the input image is taken in multiple scales (original size, r = 384 pixels, r = 256 pixels), patches are collected from all three scales simultaneously and fed into the CNN to estimate each patch's probability of being a DR lesion as in the first stage. DR maps are conducted for each image scale, maps are then averaged to produce an average probability of a positive DR prediction which ends in classifying the image into a DR stage. The authors have also finely-tuned the AlexNet classifier by replacing its last 1000- fully-connected layer with a 2-way FC layer and reinitializing its filter weights according to a gaussian distribution with std = 0.01. Stated modifications were applied due to the misselection of patches when employing default initialization of the pre-trained CNN. No dropout layer was used with the final softmax layer in the CNN. This model was evaluated on the public MESSIDOR dataset and reached an F1-score of 92.4%, a sensitivity of 99.5%, precision of 86.3%, and 0.925 AUC.

Training GoogLeNet and AlexNet CNNs for 2-ary, 3-ary, and 4-ary accurate DR grading was proposed by Lam et al. [17]. These models addressed previous limitations of identifying early stages of the disease as reported in Gargeya and Leng's system. Various image preprocessing and data augmentation methods were employed, as well as training multi-class models for the purpose of improving the algorithm's sensitivity to early-stage DR in fundal images. The deep layered CNNs were performing efficiently using a combination of heterogeneous sized filters and low-dimensional embeddings, this certainly assesses the model to learn deeper features from the training datasets. Furthermore, data augmentation methodologies used; such as image padding, rolling, rotation and zooming, were most effective in detecting R1 stage images, which were previously determined as the most difficult to classify among the 4 DR stages. In addition to applying Contrast Limited Adaptive Histogram Equalization (CLAHE) filtering algorithm, which serves the issue of retinal screening brightness and resolution that affect the intensity values of the image pixels, and therefore makes it harder to identify features to be extracted from each image (see Fig. 1). With CLAHE applied, the 3-ary classifier (None, Mild, and severe case classifier) sensitivity to identify R1/Mild cases has significantly increased from 0% as reported previously up to 29.7% when tested on the public MESSIDOR dataset. However, these models were unable to enhance the identification of multi-class DR as other previously developed systems have achieved greater sensitivity and specificity levels. Nevertheless, these experiments confirmed bright possibilities of more accurate levels for early DR detection with very subtle features, by improving image preprocessing for delicate feature extraction.

Sengupta et al. [19] introduced architecture mainly aimed to achieve a robust classification for unprecedented and varying datasets. They trained their model using Kaggle/EYEPACS dataset, which contains 5 DR stages; healthy, mild, moderate, severe, and PDR. The dataset was divided into two groups; low and high disease grades, each containing grades 1, 2 and grades 3, 4, 5 respectively. The model includes image preprocessing and augmentation, however, slightly different than Lam et al. proposed model [17], the authors rescaled image pixels, set image mean to 0 and then normalized them. Moreover, the Hough transform, which is a feature extraction technique used in image processing and computer visualization, was used whenever an image notch was identified after conversion to grayscale. Histogram Equalization was applied to all images. This was followed by a modified inception v3, 256-neuron dense layer, and a softmax layer consecutively, with 2 output probabilities of the 2 classifiers. The proposed model was tested on the MESSIDOR dataset and outperformed several previous models in grading different DR stages. It has achieved 90.4% accuracy, 89.26% sensitivity, 91.94% specificity, and an AUC of 0.90.

Due to the lack of large ground-truth-labeled datasets available for retinal funduscopy, high complexity of traditional image analysis techniques when used for large data amounts, and the quite limited models' robustness to various databases, Zago et al. [20] introduced a unique, state-of-the-art methodology for DR detection, constructed by patch-based DCNNs, this is to distinguish most challenging patches to be lesion or lesion-free patches, referred to as Red Lesions, which are classified after training the model to localize critical patches, each defined by a 65-pixel subsample. Patch localization was proven to speed-up image processing than specifically using classical image segmentation. The suggested approach was to choose only specific critical patches from each image using strides (image subsampling), instead of segmenting images and including all present patches; this, on 512 x 512-pixel images, reduces the number of patches obtained, which in turn states the number of corresponding predictions outputs from 1,300,720 to 52,428 when using stride 5 (explained in Fig. 2). In terms of time, the lesion-localization process accelerated the process from 20 minutes to only 48 seconds per image on GPU GeForce GTX 1050Ti. The proposed model achieved a sensitivity of 90%, specificity of 87%, and 0.944 AUC, implying great simplification possibilities for image processing techniques, which dramatically decreases the model's complexity as proved by the presented method.

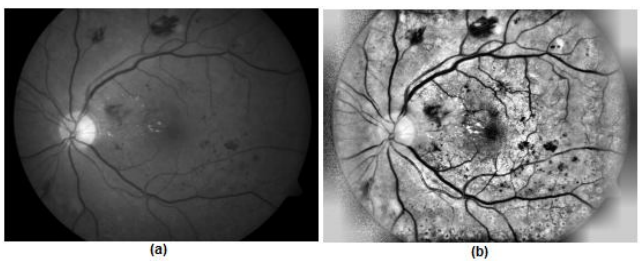

Fig. 1 - Fundus image (a) before and (b) after applying CLAHE [18].

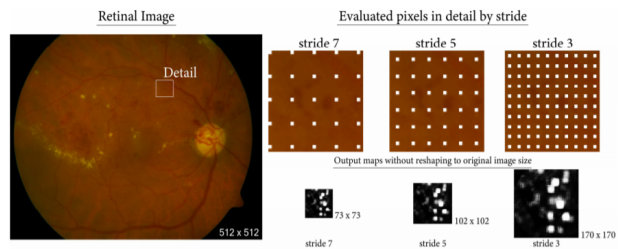

Fig. 2 - Subsampling strides technique for critical regions [20].

The CLAHE image enhancement algorithm is one of the most effective methods for image analysis, it is used for improving image quality and increasing the visibility of deep and unclear features. In addition to employing HE or CLAHE once in the prediction system as done by Lam et al. [17] and Sengupta et al. [19], Hemanth et al. [21] proved upgraded results when applying both HE and CLAHE to the dataset images. Each image in the dataset was first resized to 150 x 255-pixel, converted to RGB format; and segregated into 3 images, Red, Green, and Blue. Afterward, HE is performed upon the 3 images, followed by CLAHE and edge sharpening to reduce the noise in the image. Finally the three images are combined to produce the final improved image. A typical 8-layer CNN was used for DR classification, consisting of an input, convolutional, ReLU, cross-channel normalization, max pooling, FC, softmax and classification layers ordered respectively. The authors also used a stride function to determine the step size within the max-pooling layer through the learning process, which also downsamples the data to avoid overfitting. This present model has reached a classification accuracy of 97%, along with 94% sensitivity and 98% specificity. This study has demonstrated a significant improvement in image quality for feature extraction, when images are processed with HE and CLAHE together.

Interestingly, Shankar et al. [22] introduced an SDL-based classification model. The proposed framework consists of 3 main stages; an Input layer, a pair of Deep Convolutional Neural Networks (DCNN), and a Synergic Network (SN). A dual DCNN synergic network is designed to recognize what an image represents, providing a synergic signal that indicates whether the pair of input images belongs to the same expected input category, with a suggested solution in case an error signal occurred [23]. Image preprocessing techniques followed were initially to convert the retinal image to RGB, then applying histogram-based segmentation to extract most informative parts of the images that indicate disease-related features. A pair of preprocessed images are input to the model concurrently, each is processed through one of the DCNNs for deep feature learning, the output for each image along with its synergic signal are forwarded to the SN. as the synergic network receives 2 feature vectors from the prior DCNN phase, it concatenates, and forwards them further to a learning layer, which ends in classifying the images as one of 4 predefined

DR stages (normal, S-1, S-2, S-3). The introduced model has achieved excellent unprecedented results on the MESSIDOR dataset with 99.3% accuracy, 98% sensitivity, and 99% specificity.

Unlike previous combinations of image preprocessing techniques, Pour *et al*. [24] implemented an image-augmentation-free method for analyzing funduscopy data. The authors only employed CLAHE to enhance image quality by increasing color contrast, this method is used to determine vessel or non-vessel lesions. To increase the model's accuracy in comparison to Zago *et al*. [20], the pre-trained CNNs family, EfficientNets, was chosen to be the most effective for this diagnostic model. Moreover, pre-trained CNNs were found to save time and improve accuracy. The authors suggested scaling up CNNs to achieve greater accuracy rates, which can be performed on three dimensions; CNN's width, depth and resolution; where width implies the number of channels in each convolutional layer, depth indicates the number of layers in the network, and the resolution is determined by the resolution of the image fed to the CNN. Pour *et al*. performed a full dimensional scaling to EfficientNet-B5, which uses 456x456 pixel images. All previous models in the family (B0-B4) use smaller images which results in a loss of vital information for the classification process. The authors used Tan and Le's [25] recommended parameters for scaling the network; width =1.6 and depth = 2.2. This model was trained on a combination of datasets (MESSIDOR-2 and IDRiD) and tested on MESSIDOR, upon which it has achieved a sensitivity rate of 92% and 0.945 AUC.

### III. PROPOSED METHODOLOGY

Numerous approaches have been studied and tested in the last two decades, various image processing algorithms, classification techniques and methodologies have been used for the purpose of building a reliable automated diagnostic system for Diabetic Retinopathy. In this study, developed a state-of-the-art classification model using several ensemble learning techniques in order to attain an optimal classifier with higher precision and accuracy than previously built models. For this study, feature extraction methods and image analysis will not be processed, this classifying model will be trained on previously extracted features from the MESSIDOR public fundus dataset [27].

In this model, we introduce a promising solution for automated DR detection systems, as we will demonstrate the results of using several classification algorithms in an ensemble-based architecture. As briefed in Fig. 3, specific features will be selected and grouped using feature selection algorithms, then fed into the ensemble framework. During which, the model will be trained using part of the chosen datasets. Several widely common classification algorithms will be employed to improve the system's prediction accuracy, such as Random Forest (RF) for robust and powerful learning, Neural Network (NN) to perform complex mathematical calculations for improved precision, and a Support Vector Machine (SVM) algorithm for generating predictions in an accurate, time-saving and storage efficient manner. In the final stage, the output of all algorithms will be merged by a Meta-classifier to produce the final prediction.

#### A. Experimental Dataset

The dataset from UCI (University of California Irvine) for Diabetic Retinopathy is used in this study [12,28]. Features of this dataset have been extracted from the publicly available MESSIDOR database of 1151 fundus images of patients [29]. The model will typically go through two primary stages; training and evaluation stages. The MESSIDOR dataset will be divided to be used in both stages using the cross-validation technique. This dataset contains 1151 records of healthy (defined as 0) and multi-staged DR (defined as 1) cases, representing 540 healthy records and 611 diseased. The dataset is formed of 20 attributes. The features included were found to have a varying effect on the final prediction, therefore, feature histograms were prepared for visualization and to further assessment in the next stages of the model.

#### B. Feature Selection

In the proposed model, features are to be selected from the original dataset in the following approach: first 5 and 10 most effective features are selected to form 2 new subdatasets. The feature selection process can be performed in several ways such as Information Gain Attribute Evaluation and Wrapper Subset Evaluation which will be used in this model. Information Gain Attribute Evaluation; also known as entropy evaluation, is an attribute evaluation method which calculates the amount of information an attribute provides by giving it a value between 0 (no info.) to 1 (max info.). Most informative features are selected for further processing in the classification model while features with less valuable information are neglected. Wrapper Subset Evaluation; this feature selection method tests several subsets of features from the original dataset by a fast-learning yet powerful pre-defined algorithm, pointing out which subdataset performed better than the others, and accordingly it is selected as the most effective subdataset. This subdataset is considered to contain the most informative attributes of the original dataset.

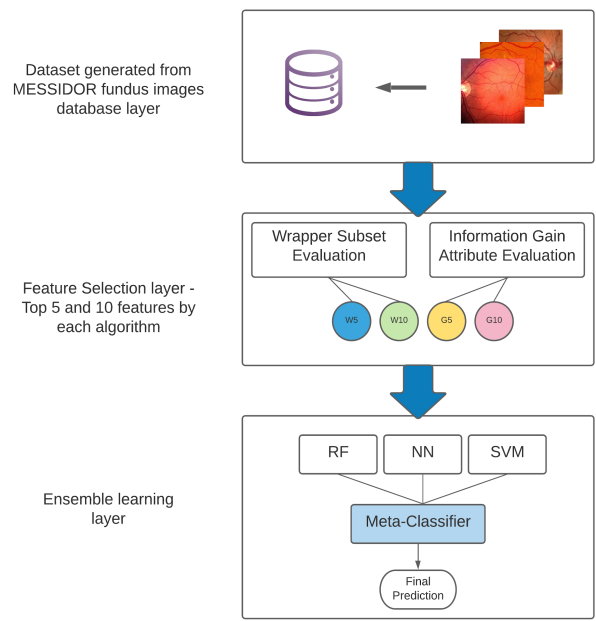

Fig. 3 - The layered model of the proposed DR detection framework.

## C. Evaluation Metrics

The performance of the proposed model was measured using a set of well-known parameters such as accuracy, precision, and recall. The classifiers' performance was measured based on the confusion matrix as follows. The true-positive (TP) rate indicates the rate of the correct predictions of the diseased cases. False-positive (FP) rate indicates the proportion of mispredicted healthy cases. The true-negative (TN) rate is an indication of the total number of healthy cases that were correctly predicted as healthy, where the false-negative (FN) rate shows the total number of DR cases that were misclassified as healthy. The precision rate represents the ratio of the total correct predictions of the DR cases to the total count of healthy and diseased patients. The recall accuracy rate represents the attribution of the correct prediction rate of the presence of the disease to the total count of DR cases. Finally the accuracy rate takes all confusion matrix parameters into its calculation to measure the correctly classified DR/Non-DR instances. Precision, recall and accuracy formulas are shown below respectively [30,31].

$$\text{Precision} = \frac{TP}{TP+FP} \quad (1)$$

$$\text{Recall} = \frac{TP}{TP+FN} \quad (2)$$

$$\text{Accuracy} = \frac{TP+TN}{TP+TN+FP+FN} \quad (3)$$

The F-Measure metric was measured in this research paper, where its calculation depends on the following formula.

$$\text{F-Measure} = 2 \times \frac{Precision \times Recall}{Precision+Recall} \quad (4)$$

## D. Ensemble learning

Ensemble systems, also known as multiple-classifier systems, are obtained by combining the knowledge gained by multiple contributing models to form one final decision [32]. In the last few years, ensembles have become more popular in use for decision-making systems, as a consequence of their great performance when run for prediction, classification, and regression problems. This approach of knowledge accumulation enhances the overall decision of the system as a whole, and each of the models individually as well. There are two main reasons to use an ensemble-based over a single-based model, Performance, and Robustness; an ensemble can make better predictions and achieve better performance than any single contributing model. In addition, an ensemble reduces the spread or dispersion of the predictions and model performance. In this model, 3 main classification algorithms were embedded in the ensemble, Random Forest, Neural Network, and Support Vector Machine.

## E. Results and Discussion

The evaluation process can be divided into 3 stages. Firstly, the model along with single classifiers were evaluated to extract the best single classifier and the best performing subdataset, Neural Networks and InfoGain top 5. Secondly, the best single classifier is compared to the proposed ensemble when evaluated on the Messidor dataset, the proposed ensemble overperformed the Neural Networks in this step. Finally, the present ensemble was tested on all 4 subdatasets, which reassured the importance of the InfoGainEval. top 5 features. Table 1 shows the obtained accuracy for each of the contributing datasets in evaluating the proposed model, indicating as well the performance of the ensemble when based on a stacking architecture. In Table 2, a comparison is presented between the proposed ensemble and the best single classifier on the Messidor dataset in terms of accuracy, recall, precision, and AUC. In addition, Table 3 extensively compares the performance of the selected sets of features, which implies the impressive capability of the InfoGainEval. top 5 features. Moreover, Figure 4 presents the accuracy rates achieved for the best performing subdataset by each contributing classifier against the proposed stacking ensemble. The outcomes of assembling multiple classification algorithms such as Random Forest, Neural Networks, and Support Vector Machine in a stacking framework implied improved final predictions. Furthermore, previous models were trained and tested after performing several image processing and enhancing algorithms on the original fundus images dataset for feature extraction, while in this model it was completely trained on MESSIDOR's previously-extracted features. Image processing and analysis were not carried out in this study, it was a real challenge to achieve high accuracy rates with a very limited number of features.

Table 1 - Accuracy rates of single classifiers and the proposed ensemble. *mean(std)

|  | *SVM* | *NN* | *RF* | *Proposed* |
|---|---|---|---|---|
| **Original dataset** | 0.697(0.041) | 0.719(0.041) | 0.686(0.032) | **0.751(0.033)** |
| **Wrapper top 5** | 0.553(0.022) | 0.575(0.042) | 0.581(0.028) | **0.588(0.044)** |
| **Wrapper top 10** | 0.582(0.031) | 0.618(0.025) | 0.634(0.046) | **0.645(0.044)** |
| **InfoGain top 5** | 0.685(0.050) | 0.696(0.053) | 0.670(0.034) | **0.707(0.031)** |
| **InfoGain top 10** | 0.685(0.046) | 0.674(0.045) | 0.670(0.035) | **0.701(0.031)** |

Table 2 - Best single classifier and the proposed ensemble performance evaluated on Messidor dataset.

| *Model* | *Accuracy* | *Recall* | *Precision* | *AUC* |
|---|---|---|---|---|
| **Best Single Classifier** | 0.719 | 0.704 | 0.774 | 0.816 |
| **Proposed Model** | 0.751 | 0.725 | 0.778 | 0.827 |

Table 3 - Proposed model evaluation on selected features of Messidor dataset.

| *Subdataset* | *Accuracy* | *Recall* | *Precision* | *AUC* |
|---|---|---|---|---|
| **Wrapper top 5** | 0.588 | 0.512 | 0.635 | 0.619 |
| **Wrapper top 10** | 0.645 | 0.656 | 0.677 | 0.701 |
| **InfoGain top 5** | **0.707** | **0.709** | **0.763** | **0.782** |
| **InfoGain top 10** | 0.701 | 0.679 | 0.741 | 0.770 |

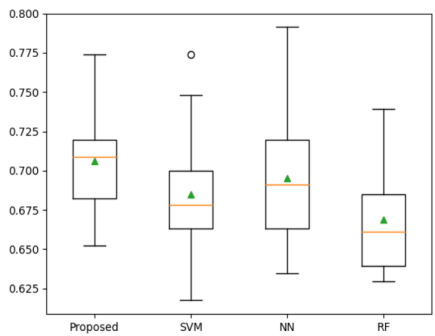

Figure 4 - Classifiers' accuracy on the best performing subdataset InfoGain top 5.

## IV. CONCLUSION

This research points out promising technological advancements for the healthcare and medical sectors, especially in the early detection of many types of illnesses. Each and every disease is best treated when in its earliest stages, such as, and most importantly, Cancer, Diabetic Retinopathy, Cholesterol abnormalities, and many others. Moreover, automatic detection models are time and cost-efficient, which will serve various communities and regions, and can be run by any practitioner once they are familiar with the models processing and how decisions are displayed. In the present work, we introduce a new framework for Diabetic Retinopathy detection using Ensemble Machine Learning. In addition, in order to fill the gap of insufficient performance in the previous models, we have applied several feature engineering techniques as well as a substantial stack of classification algorithms, with a final Meta-Classifier, that is to ensure acquiring the highest accuracy while implementing a reliable and robust diagnostic model for different patients around the world. It is to be noted that the current results are initial indicators, other ensembles are being implemented presently to upgrade the overall performance of the model.